\documentstyle[11pt,amssymb,epsfig,psfig]{article}

\begin{document}

\title{{\bf Casimir densities for two concentric spherical shells in the global
monopole spacetime }}

\author{A. A. Saharian  $^{1}$\footnote{E-mail:
saharyan@www.physdep.r.am }
and  M. R. Setare $^2$ \footnote{E-mail: rezakord@yahoo.com} \\
 {$^1$ Department of Physics, Yerevan
State University, Yerevan, Armenia } \\
and \\
{$^2$ Institute for Theoretical Physics and Mathematics,
Tehran, Iran} \\
{Department of Science, Physics Group, Kurdistan
University, Sanandeg, Iran}\\
{Department of Physics, Sharif
University of Technology, Tehran, Iran}}

\maketitle

\begin{abstract}
The quantum vacuum effects are investigated for a massive scalar
field with general curvature coupling and obeying the Robin
boundary conditions given on two concentric spherical shells with
radii $a $ and $b$ in the $D+1$-dimensional global monopole
background. The expressions are derived for the Wightman function,
the vacuum expectation values of the field square, the vacuum
energy density, radial and azimuthal stress components in the
region between the shells. A regularization procedure is carried
out by making use of the generalized Abel-Plana formula for the
series over zeros of combinations of the cylinder functions. This
formula allows us to extract from the vacuum expectation values
the parts due to a single sphere on background of the global
monopole gravitational field, and to present the "interference"
parts in terms of exponentially convergent integrals, useful, in
particular, for numerical evaluations. The vacuum forces acting on
the boundaries are presented as a sum of the self--action and
interaction terms. The first one contains well known surface
divergences and needs a further regularization. The interaction
forces between the spheres are finite for all values $a<b$ and are
attractive for a Dirichlet scalar. The asymptotic behavior of the
vacuum densities is investigated (i) in the limits $a\to 0$ and
$b\to \infty $, (ii) in the limit $a,b\to \infty $ for fixed value
$b-a$, and (iii) for small values of the parameter associated with
the solid angle deficit in global monopole geometry. We show that
in the case (ii) the results for two parallel Robin plates on the
Minkowski bulk are rederived to the leading order.
\end{abstract}

\bigskip

{PACS number(s): 03.70.+k, 11.10.Kk}

\newpage

\section{Introduction}

The Casimir effect is regarded as one of the most striking
manifestation of vacuum fluctuations in quantum field theory. The
presence of reflecting boundaries alters the zero-point modes of a
quantized field, and results in the shifts in the vacuum
expectation values of quantities quadratic in the field, such as
the energy density and stresses. In particular, vacuum forces
arise acting on constraining boundaries. The particular features
of these forces depend on the nature of the quantum field, the
type of spacetime manifold and its dimensionality, the boundary
geometries and the specific boundary conditions imposed on the
field. Since the original work by Casimir in 1948 \cite{Casi48}
many theoretical and experimental works have been done on this
problem (see, e.g.,
\cite{Most97,Plun86,Lamo99,Bord99,Bord01,Kirs01,Bord02,Milt02} and
references therein). Many different approaches have been used:
mode summation method with combination of the zeta function
regularization technique, Green function formalism, multiple
scattering expansions, heat-kernel series, etc. The Casimir effect
can be viewed as a polarization of vacuum by boundary conditions.
Another type of vacuum polarization arises in the case of external
gravitational fields \cite{Birr82,Grib94}. In a previous paper
\cite{Saha03} we have studied an example of a situation when both
types of sources for the polarization are present. Namely, we have
investigated the vacuum expectation values of the square of a
scalar field and energy-momentum tensor induced by a spherical
shell in the spacetime of a point-like global monopole.

Topological defects have attracted a great deal of attention
because of their relevance to a number of different areas ranging
from condensed matter to structure formation (for a review see
\cite{Vile94}). In the context of hot big bang cosmology, the
unified theories of the fundamental interactions predict that the
universe passes through a sequence of phase transitions. These
phase transitions might have given rise to several kinds of
topological defects depending on the nature of the symmetry that
is broken \cite{Kibb76}. If a global SO(3) symmetry of a triplet
scalar field is broken, the point like defects called global
monopoles are believed to be formed. The simplified global
monopole was introduced by Sokolov and Starobinsky \cite{Soko77}.
The gravitational effects of the global monopole were studied in
Ref. \cite{Barr89}, where a solution is presented which describes
a global monopole at large radial distances. The quantum vacuum
effects of the matter fields on the global monopole background
have been considered in \cite{Hisc90,Mazz91,Beze99,Beze02}. The
effects produced by the non-zero temperature are investigated as
well \cite{Beze01temp}. The zeta function and the Casimir energy
for spherical boundaries in this background are considered in
\cite{Bord96m,Beze00}.

In this paper, we continue the investigation started in
\cite{Saha03} and consider the Casimir densities in the region
between two concentric spherical shells on background of the
$D+1$-dimensional spacetime of a point-like global monopole. The
positive frequency Wightman function, the vacuum expectation
values of the field square and energy-momentum tensor are
investigated for a massive scalar field with general curvature
coupling parameter $\xi $. In addition to describing the physical
structure of the quantum field at a given point, the
energy-momentum tensor acts as the source of gravity in the
Einstein equations. It therefore plays an important role in
modeling a self-consistent dynamics involving the gravitational
field \cite{Birr82}. We study the general case of Robin boundary
conditions with different coefficients for the inner and outer
spheres. The Robin boundary conditions are an extension of the
ones imposed on perfectly conducting boundaries and may, in some
geometries, be useful for depicting the finite penetration of the
field into the boundary with the "skin-depth" parameter related to
the Robin coefficient \cite{Mostep,Lebedev}. It is interesting to
note that the quantum scalar field satisfying Robin condition on
the boundary of a cavity violates the Bekenstein's
entropy-to-energy bound near certain points in the space of the
parameter defining the boundary condition \cite{Solod}. This type
of conditions also appear in considerations of the vacuum effects
for a confined charged scalar field in external fields
\cite{Ambjorn2} and in quantum gravity \cite{Mo,EK,Esp97}. Mixed
boundary conditions naturally arise for scalar and fermion bulk
fields in the Randall-Sundrum model \cite{Gherg}. In this model
the bulk geometry is a slice of anti-de Sitter space and the
corresponding Robin coefficient is related to the curvature scale
of this space. For scalars with general curvature coupling the
essential point is the relation between the mode sum energy,
evaluated as a renormalized sum of the zero-point energies for
each normal mode of frequency, and the volume integral of the
renormalized energy density. In \cite{Rome02} it has been shown
that in the discussion of this question for the Robin parallel
plates it is necessary to include in the energy a surface term
concentrated on the boundary (see \cite{Saha01,Rome01,Saha02} for
similar issues in more complicated geometries and the discussion
of the paper \cite{Rome02} in \cite{Full03}).

We have organized the paper as follows. In the next section we
derive a formula for the Wightman function in the region between
two spheres. The reason for our choice of the Wightman function is
that this function also determines the response of the particle
detectors in a given state of motion. Following Refs.
\cite{Saha01,Rome01,Saha02,Avag02}, to evaluate the bilinear field
products we use the mode sum method in combination with the
summation formulae from \cite{Saha87} (see also \cite{Saha00rev}).
These formulae allow (i) to extract from vacuum expectation values
the parts due to a single sphere on the global monopole
background, and (ii) to present the "interference" parts in terms
of exponentially convergent integrals involving the modified
Bessel functions. The vacuum expectation value of the field square
is obtained computing the Wightman function in the coincidence
limit, and is investigated in section \ref{secm:3}. The various
limiting cases are considered including the limit of the strong
gravitational field corresponding to small values of the parameter
associated with the solid angle deficit in the global monopole
spacetime. Section \ref{secm:4} is devoted to the vacuum
expectation values of the energy density and stresses. The
formulae are derived for the interaction forces between the
spheres. The behavior of these quantities is investigated in
various limiting cases. Section \ref{secm:5} concludes the main
results of the paper.

\section{Wightman function } \label{secm:2}

In this paper we consider a real scalar field $\varphi $ with
curvature coupling parameter $\xi $ in the $D+1$-dimensional
spacetime of a point-like global monopole. In the hyperspherical
polar coordinates $(r,\vartheta ,\phi )\equiv (r,\theta
_{1},\theta _{2},\ldots \theta _{n},\phi )$, $n=D-2$, the
corresponding geometry is described by the line element
\begin{equation}
ds^{2}=dt^{2}-dr^{2}-\sigma ^{2}r^{2}d\Omega _{D}^{2},  \label{mmetric}
\end{equation}
where $d\Omega _{D}^{2}$ is the line element on the surface of the
unit sphere in $D$-dimensional Euclidean space, the parameter
$\sigma $ is smaller than unity and is related to the symmetry
breaking energy scale in the theory. In the spacetime given by
line element (\ref{mmetric}) the solid angle deficit is $(1-\sigma
^{2})S_{D}$, with $S_{D}=2\pi ^{D/2}/\Gamma (D/2)$ being the total
area of the surface of the unit sphere in $D$-dimensional
Euclidean space. The field equation has the form
\begin{equation}
\left( \nabla _{i}\nabla ^{i}+m^{2}+\xi R\right) \varphi =0,\quad
\label{mfieldeq}
\end{equation}
where $m$ is the mass for the field quanta, $\nabla _{i}$ is the
covariant derivative operator associated with the metric given by
line element (\ref{mmetric}), and
\begin{equation}\label{Riccisc}
  R=n(n+1)\frac{\sigma ^{2}-1}{\sigma ^{2}r^{2}}
\end{equation}
is the corresponding Ricci scalar (we adopt the convention of
Birrell and Davies \cite{Birr82} for the curvature tensor). Note
that for $\sigma \neq 1$ the geometry is singular at the origin
(point-like monopole), $r=0$. In (\ref{mfieldeq}) the values of
the curvature coupling parameter $\xi =0$, and $ \xi =\xi _{D}$
with $\xi _{D}\equiv (D-1)/4D$ correspond to the minimal and
conformal couplings, respectively.

In this paper we are interested in the vacuum expectation values
(VEVs) of the field bilinear products on background of the
geometry described by (\ref{mmetric}), assuming that the field
satisfies the Robin boundary conditions
\begin{equation}
\left( \tilde{A}_{r}+\tilde{B}_{r}\frac{\partial }{\partial r}\right)
\varphi (x)=0,\quad r=a,b,  \label{mrobcond}
\end{equation}
on two spheres with radii $a$ and $b$, $a<b$, concentric with the
monopole. Here the coefficients $\tilde{A}_{r}$ and
$\tilde{B}_{r}$ are constants, in general, different for the inner
and outer spheres. The imposition of this boundary condition on
the quantum field $\varphi (x)$ leads to the modification of the
spectrum for the zero-point fluctuations and results in the shift
in VEVs for physical quantities. In particular, vacuum forces
arise acting on constraining boundary. This is the familiar
Casimir effect. The VEVs for the physical  quantities bilinear in
the field can be evaluated if the corresponding Wightman function
is known. For this reason we first concentrate on the positive
frequency Wightman function $ G^{+}(x,x^{\prime })=\langle
0|\varphi (x)\varphi (x^{\prime })|0\rangle $, where $|0\rangle $
is the amplitude for the corresponding vacuum state. The Wightman
function also determines the response of the particle detectors in
a given state of motion \cite{Birr82}. By expanding the field
operator over eigenfunctions and using the commutation rules one
can see that
\begin{equation}
\langle 0|\varphi (x)\varphi (x^{\prime })|0\rangle =\sum_{\alpha }\varphi
_{\alpha }(x)\varphi _{\alpha }^{\ast }(x^{\prime }),  \label{mfieldmodesum}
\end{equation}
where $\left\{ \varphi _{\alpha }(x),\varphi _{\alpha }^{\ast
}(x^{\prime })\right\} $ is a complete orthonormal set of positive
and negative frequency solutions to the field equation with
quantum numbers $\alpha $, satisfying boundary conditions
(\ref{mrobcond}). Note that for $D=1$ we have the standard
Casimir-like geometry on Minkowski spacetime.

In the hyperspherical coordinates, for the region between two
spheres the complete set of solutions to (\ref{mfieldeq}) with
scalar curvature from (\ref{Riccisc}), has the form
\begin{equation}
\varphi _{\alpha }(x)=\beta _{\alpha }r^{-n/2}g_{\nu _{l}}(\lambda a,\lambda
r)Y(m_{k};\vartheta ,\phi )e^{-i\omega t},\,\,l=0,1,2,\ldots ,
\label{meigfunc}
\end{equation}
where $m_{k}=(m_{0}\equiv l,m_{1},\ldots ,m_{n})$, and $m_{1},m_{2},\ldots
,m_{n}$ are integers such that
\begin{equation}
0\leq m_{n-1}\leq m_{n-2}\leq \cdots \leq m_{1}\leq l,\quad
-m_{n-1}\leq m_{n}\leq m_{n-1},  \label{mnumbvalues}
\end{equation}
$Y(m_{k};\vartheta ,\phi )$ is the surface harmonic of degree $l$
(see \cite {Erdelyi}). In (\ref{meigfunc})
\begin{equation}
g_{\nu _{l}}(\lambda a,\lambda r)\equiv J_{\nu _{l}}(\lambda r)\bar{Y}_{\nu
_{l}}^{(a)}(\lambda a)-\bar{J}_{\nu _{l}}^{(a)}(\lambda a)Y_{\nu
_{l}}(\lambda r),\quad \lambda =\sqrt{\omega ^{2}-m^{2}},  \label{genu}
\end{equation}
$J_{\nu }(z)$ and $Y_{\nu }(z)$ are the Bessel and Neumann
functions, and the functions with overbars are defined in
accordance with
\begin{equation}
\bar{F}^{(\alpha )}(z)\equiv A_{\alpha }F(z)+B_{\alpha }zF^{\prime
}(z),\quad A_{\alpha }=\tilde{A}_{\alpha }-\tilde{B}_{\alpha }n/2\alpha
,\quad B_{\alpha }=\tilde{B}_{\alpha }/\alpha ,\;\alpha =a,b.
\label{barnotab}
\end{equation}
Here and below we use the following notation
\begin{equation}
\nu _{l} =\frac{1}{\sigma }\left[ \left( l+\frac{n}{2}\right)
^{2}+(1-\sigma ^{2})n(n+1)\left( \xi -\xi _{D-1}\right) \right]
^{1/2}, \label{mnulam}
\end{equation}
assuming that $\nu _{l}^{2}$ is non-negative. This corresponds to the
restriction on the values of the curvature coupling parameter for $n>0$,
given by the condition
\begin{equation}
\xi \geq -\frac{n\sigma ^{2}}{4(n+1)(1-\sigma ^{2})}.
\end{equation}
This condition is satisfied by the most important special cases of
the minimal and conformal couplings. The coefficients $\beta
_{\alpha }$ in (\ref{meigfunc}) can be found from the
normalization condition
\begin{equation}
\int \left| \varphi _{\alpha }(x)\right| ^{2}\sqrt{-g}dV=\frac{1}{2\omega },
\label{normcond}
\end{equation}
where the integration goes over the region between the spheres, $a\leq r\leq
b$. Substituting eigenfunctions (\ref{meigfunc}), and using the relation
\begin{equation}
\int \left| Y(m_{k};\vartheta ,\phi )\right| ^{2}d\Omega =N(m_{k})
\end{equation}
(the explicit form for $N(m_{k})$ is given in \cite{Erdelyi} and
will not be necessary for the following considerations in this
paper) for the spherical harmonics and the value for the standard
integral involving the square of a cylinder function
\cite{Prudnikov}, one finds
\begin{equation}
\beta _{\alpha }^{2}=\frac{\pi ^{2}\lambda T_{\nu _{l}}^{ab}(b/a,\lambda a)}{%
4N(m_{k})\omega a\sigma ^{D-1}},  \label{normcoef}
\end{equation}
where we use the notation
\begin{equation}
T_{\nu }^{ab}(\eta ,z)=z\left\{ \frac{\bar{J}_{\nu }^{(a)2}(z)}{\bar{J}_{\nu
}^{(b)2}(\eta z)}\left[ A_{b}^{2}+B_{b}^{2}(\eta ^{2}z^{2}-\nu ^{2})\right]
-A_{a}^{2}-B_{a}^{2}(z^{2}-\nu ^{2})\right\} ^{-1},\quad \eta =\frac{b}{a}.
\label{tekaAB}
\end{equation}
The functions chosen in the form (\ref{genu}) satisfy the boundary
condition on the sphere $r=a$. From the boundary condition on
$r=b$ one obtains that the corresponding eigenmodes are solutions
to the equation
\begin{equation}
C_{\nu _{l}}^{ab}(b/a,\lambda a)\equiv \bar{J}_{\nu }^{(a)}(\lambda a)\bar{Y}%
_{\nu }^{(b)}(\lambda b)-\bar{J}_{\nu }^{(b)}(\lambda b)\bar{Y}_{\nu
}^{(a)}(\lambda a)=0.  \label{eigmodesab}
\end{equation}
Below the roots to this equation will be denoted by $\gamma _{\nu
,k}=\lambda a$ , $k=1,2,\ldots $. The corresponding eigenfrequencies $\omega
=\omega _{\nu _{l},k}$ are related to these zeros as $\omega _{\nu _{l},k}=%
\sqrt{\lambda _{\nu _{l},k}^{2}/a^{2}+m^{2}}$.

Substituting the eigenfunctions into the mode sum
(\ref{mfieldmodesum}) and using the addition formula
\begin{equation}
\sum_{m_{k}}\frac{1}{N(m_{k})}Y(m_{k};\vartheta , \phi
)Y^{*}(m_{k};\vartheta ^{\prime },\phi ^{\prime
})=\frac{2l+n}{nS_{D}}C_{l}^{n/2}(\cos \theta ), \label{adtheorem}
\end{equation}
for the expectation value of the field product one finds
\begin{equation}
\langle 0|\varphi (x)\varphi (x^{\prime })|0\rangle =\frac{\pi
^{2}(rr^{\prime })^{-n/2}}{4naS_{D}\sigma ^{D-1}}\sum_{l=0}^{\infty
}(2l+n)C_{l}^{n/2}(\cos \theta )\sum_{k=1}^{\infty }h(\gamma _{\nu
_{l},k})T_{\nu _{l}}^{ab}(b/a,\gamma _{\nu _{l},k}),  \label{fieldmodesum1ab}
\end{equation}
with the function
\begin{equation}
h(z)=\frac{z}{\sqrt{z^{2}+m^{2}a^{2}}}g_{\nu _{l}}(z,zr/a)g_{\nu
_{l}}(z,zr^{\prime }/a)e^{i\sqrt{z^{2}/a^{2}+m^{2}}(t^{\prime
}-t)}, \label{hab}
\end{equation}
and $C_{l}^{p}(x)$ is the Gegenbauer or ultraspherical polynomial
of degree $l$ and order $p$. To sum over $k$ we will use the
generalized Abel-Plana summation formula
\cite{Saha87,Saha00rev,Saha01}
\begin{eqnarray}
\frac{\pi ^{2}}{2}\sum_{k=1}^{\infty }h(\gamma _{\nu ,k})T_{\nu }^{ab}(\eta
,\gamma _{\nu ,k}) &=&\int_{0}^{\infty }\frac{h(x)dx}{\bar{J}_{\nu
}^{(a)2}(x)+\bar{Y}_{\nu }^{(a)2}(x)}  \label{cor3form} \\
&&-\frac{\pi }{4}\int_{0}^{\infty }\frac{\bar{K}_{\nu }^{(b)}(\eta x)}{\bar{K%
}_{\nu }^{(a)}(x)}\frac{\left[ h(xe^{\pi i/2})+h(xe^{-\pi i/2})\right] dx}{%
\bar{K}_{\nu }^{(a)}(x)\bar{I}_{\nu }^{(b)}(\eta x)-\bar{K}_{\nu
}^{(b)}(\eta x)\bar{I}_{\nu }^{(a)}(x)},  \nonumber
\end{eqnarray}
where $I_{\nu }(x)$ and $K_{\nu }(x)$ are the Bessel modified
functions. The corresponding conditions for the function
(\ref{hab}) are satisfied if $r+r^{\prime }+|t-t^{\prime }|<2b$.
Note that this is the case in the coincidence limit for the region
under consideration. Note that we have assumed values $A_{\alpha
}$ and $B_{\alpha }$ for which all zeros for (\ref{eigmodesab})
are real. In case of existence of purely imaginary zeros we have
to include additional residue terms on the left of this formula
(see \cite{Saha87,Saha00rev,Saha01}). Applying to the sum over $k$
in (\ref{fieldmodesum1ab}) formula (\ref{cor3form}), for the
corresponding Wightman function one obtains
\begin{equation}
\langle 0|\varphi (x)\varphi (x^{\prime })|0\rangle =\frac{\sigma ^{1-D}}{%
2naS_{D}}\sum_{l=0}^{\infty }\frac{2l+n}{(rr^{\prime })^{n/2}}%
C_{l}^{n/2}(\cos \theta )\left\{ \int_{0}^{\infty }\frac{h(z)dz}{\bar{J}%
_{\nu _{l}}^{(a)2}(z)+\bar{Y}_{\nu _{l}}^{(a)2}(z)}\right.
\label{unregWightab}
\end{equation}
\[
-\left. \frac{2}{\pi }\int_{ma}^{\infty }\frac{zdz}{\sqrt{z^{2}-a^{2}m^{2}}}%
\frac{\bar{K}_{\nu _{l}}^{(b)}(\eta z)}{\bar{K}_{\nu _{l}}^{(a)}(z)}\frac{%
G_{\nu _{l}}^{(a)}(z,zr/a)G_{\nu _{l}}^{(a)}(z,zr^{\prime }/a)}{\bar{K}_{\nu
_{l}}^{(a)}(z)\bar{I}_{\nu _{l}}^{(b)}(\eta z)-\bar{K}_{\nu _{l}}^{(b)}(\eta
z)\bar{I}_{\nu _{l}}^{(a)}(z)}\cosh \left[ \sqrt{z^{2}/a^{2}-m^{2}}%
(t^{\prime }-t)\right] \right\} ,
\]
where we have introduced notations
\begin{equation}
G_{\nu }^{(\alpha )}(z,y)=I_{\nu }(y)\bar{K}_{\nu }^{(\alpha )}(z)-\bar{I}%
_{\nu }^{(\alpha )}(z)K_{\nu }(y),\;\alpha =a,b.  \label{Geab}
\end{equation}
In the limit $b\rightarrow \infty $ the second integral on the
right of (\ref{unregWightab}) tends to zero (for large $b/a$ the
subintegrand is proportional to $e^{-2bz/a}$), whereas the first
one does not depend on $b$. It follows from here that the term
with the first integral in the figure braces corresponds to the
Wightman function for the region outside a single sphere with
radius $a$ on background of the global monopole geometry. As a
result the Wightman function in the region between two spheres is
presented in the form
\begin{eqnarray}
\langle 0|\varphi (x)\varphi (x^{\prime })|0\rangle &=&\langle
0^{(a)}|\varphi (x)\varphi (x^{\prime })|0^{(a)}\rangle
-\frac{\sigma ^{1-D}}{\pi nS_{D}} \sum_{l=0}^{\infty
}\frac{2l+n}{(rr^{\prime })^{n/2}}C_{l}^{n/2}(\cos \theta
)  \label{regWightab1} \\
&\times &\int_{m}^{\infty }dz\frac{z\Omega _{a\nu _{l}}(az,b
z)}{\sqrt{z^{2}-m^{2}}}G_{\nu _{l}}^{(a)}(az,r z)G_{\nu
_{l}}^{(a)}(az,r^{\prime }z)\cosh \left[
\sqrt{z^{2}-m^{2}}(t^{\prime }-t)\right] ,  \nonumber
\end{eqnarray}
where
\begin{equation}
\Omega _{a\nu }(az,bz)=\frac{\bar{K}_{\nu }^{(b)}(bz)/\bar{K}_{\nu
}^{(a)}(az)}{\bar{K}_{\nu }^{(a)}(az)\bar{I}_{\nu }^{(b)}(bz)-\bar{K}_{\nu
}^{(b)}(bz)\bar{I}_{\nu }^{(a)}(az)},  \label{Omega}
\end{equation}
and $|0^{(a)}\rangle $ is the amplitude for the vacuum state in
the case of a single sphere with radius $a$. The Wightman function
$\langle 0^{(a)}|\varphi (x)\varphi (x^{\prime })|0^{(a)}\rangle$
is investigated in a previous paper \cite{Saha03}, and below we
will mainly concentrate on the terms induced by the presence of
the second sphere. Note that in the coincidence limit, $x^{\prime
}=x$, the second summand on the right hand side of
(\ref{regWightab1})  will give a finite result for $a\leq r<b$,
and is divergent on the boundary $r=b$. In Ref. \cite{Saha03} the
Wightman function for a single sphere in the global monopole
spacetime is presented in the form
\begin{equation}\label{Wmon1sph}
\langle 0^{(a)}|\varphi (x)\varphi (x^{\prime })|0^{(a)}\rangle =
\langle 0_{{\mathrm{m}}}|\varphi (x)\varphi (x^{\prime
})|0_{{\mathrm{m}}}\rangle +\langle \varphi (x)\varphi (x^{\prime
})\rangle _b^{(a)},
\end{equation}
where $|0_{{\mathrm{m}}}\rangle $ is the amplitude for the vacuum
state in the case of the boundary-free global monopole geometry.
The expressions for the boundary induced part $\langle \varphi
(x)\varphi (x^{\prime })\rangle _b^{(a)}$ for both regions inside
and outside a single shell are given in \cite{Saha03}.

Using the identity
\begin{eqnarray}
  \Omega _{\alpha \nu }(a z,b z)G_{\nu }^{(\alpha )}(\alpha z,r z)
  G_{\nu }^{(\alpha )}(\alpha z,r' z)|_{\alpha =a}^{\alpha =b} &=&
  \frac{\bar I_{\nu }^{(a)}(a z)}{\bar K_{\nu }^{(a)}(a z)}
  K_{\nu }(r z)K_{\nu }(r' z) \nonumber \\
  && - \frac{\bar K_{\nu }^{(b)}(b z)}{\bar I_{\nu }^{(b)}(b z)}
  I_{\nu }(r z)I_{\nu }(r' z), \label{ident1}
\end{eqnarray}
it can be seen that for the case of two spheres the Wightman
function in the intermediate region can also be presented in the
form
\begin{eqnarray}
\langle 0|\varphi (x)\varphi (x^{\prime })|0\rangle &=&\langle
0^{(b)}|\varphi (x)\varphi (x^{\prime })|0^{(b)}\rangle
-\frac{\sigma ^{1-D}}{\pi nS_{D}} \sum_{l=0}^{\infty
}\frac{2l+n}{(rr^{\prime })^{n/2}}C_{l}^{n/2}(\cos \theta
)  \label{regWightab2} \\
&\times &\int_{m}^{\infty }dz\frac{z\Omega _{b\nu _{l}}(az,b
z)}{\sqrt{z^{2}-m^{2}}}G_{\nu _{l}}^{(b)}(b z,r z)G_{\nu
_{l}}^{(b)}(b z,r^{\prime }z)\cosh \left[
\sqrt{z^{2}-m^{2}}(t^{\prime }-t)\right] ,  \nonumber
\end{eqnarray}
with $\langle 0^{(b)}|\varphi (x)\varphi (x^{\prime
})|0^{(b)}\rangle $ being the Wightman function for the vacuum
inside a single sphere with radius $b$, and
\begin{equation}
\Omega _{b\nu }(az,bz)=\frac{\bar{I}_{\nu }^{(a)}(az)/\bar{I}_{\nu
}^{(b)}(bz)}{\bar{K}_{\nu }^{(a)}(az)\bar{I}_{\nu }^{(b)}(bz)-\bar{K}_{\nu
}^{(b)}(bz)\bar{I}_{\nu }^{(a)}(az)}.  \label{Omegatilde}
\end{equation}
Note that formula (\ref{regWightab2}) can be also derived by the
procedure described above for (\ref{regWightab1}), if in
expression (\ref{meigfunc}) for the eigenfunctions we replace the
function $g_{\nu _l}(\lambda a,\lambda r)$ (given by (\ref{genu}))
by the function $J_{\nu _{l}}(\lambda r)\bar{Y}_{\nu
_{l}}^{(b)}(\lambda b)-\bar{J}_{\nu _{l}}^{(b)}(\lambda b)Y_{\nu
_{l}}(\lambda r)$. In the coincidence limit, the second summand on
the right of formula (\ref{regWightab2}) is finite for $a<r\leq b$
and diverges on the boundary $r=a$. It follows from here that if
we write the regularized Wightman function in the form
\begin{equation}
\langle 0|\varphi (x)\varphi (x^{\prime })|0\rangle =\langle
0_{{\mathrm{m}}}|\varphi (x)\varphi (x^{\prime
})|0_{{\mathrm{m}}}\rangle +\langle \varphi (x)\varphi (x^{\prime
})\rangle _b^{(a)}+\langle \varphi (x)\varphi (x^{\prime })\rangle
_b^{(b)}+\Delta W(x,x^{\prime }), \label{intWeight}
\end{equation}
then in the coincidence limit the ''interference'' term $\Delta
W(x,x^{\prime })$ is finite for all values $a\leq r\leq b$. Using
the formula for the Wightman function for a single sphere (see
\cite{Saha03}) and equation (\ref{regWightab1}), it can be seen
that this term may be presented as
\begin{equation}
\Delta W(x,x^{\prime })=\frac{-\sigma ^{1-D}}{\pi nS_{D}}\sum_{l=0}^{\infty }%
\frac{2l+n}{(rr^{\prime })^{n/2}}C_{l}^{n/2}(\cos \theta )\int_{m}^{\infty }%
\frac{zdz}{\sqrt{z^{2}-m^{2}}}W^{(ab)}(r,r^{\prime })\cosh \left[ \sqrt{%
z^{2}-m^{2}}(t^{\prime }-t)\right] ,  \label{intWeightrr}
\end{equation}
where
\begin{equation}
W^{(ab)}(r,r^{\prime })=\frac{\bar{I}_{\nu _{l}}^{(a)}(az)}{\bar{I}_{\nu
_{l}}^{(b)}(bz)}\frac{\bar{K}_{\nu _{l}}^{(b)}(bz)}{\bar{K}_{\nu
_{l}}^{(a)}(az)}\left[ \frac{G_{\nu _{l}}^{(a)}(az,zr)G_{\nu
_{l}}^{(b)}(bz,zr^{\prime })}{\bar{K}_{\nu _{l}}^{(a)}(az)\bar{I}_{\nu
_{l}}^{(b)}(bz)-\bar{K}_{\nu _{l}}^{(b)}(bz)\bar{I}_{\nu _{l}}^{(a)}(az)}%
-I_{\nu _{l}}(zr^{\prime })K_{\nu _{l}}(zr)\right] .  \label{Wrr}
\end{equation}
Formula (\ref{intWeight}) presents the Wightman function in the
region between two spheres, $a\leq r\leq b$. In the regions $r\leq
a $ and $r\geq b$, the Wightman functions are given by the
formulae corresponding to a single sphere with radius $a$ and $b$
respectively, and are investigated in \cite{Saha03}.

\section{Vacuum expectation value for the field square}
\label{secm:3}

The VEV of the field square is obtained computing the Wightman
function in the coincidence limit $x^{\prime }\rightarrow x$. In
the region between the spheres, from (\ref{regWightab1}) and
(\ref{regWightab2}) we obtain two equivalent forms
\begin{equation}\label{fieldsqform1}
  \langle 0|\varphi ^{2}(x)|0\rangle =\langle 0^{(\alpha )}|
  \varphi ^{2}(x)|0^{(\alpha )}\rangle
  -\frac{\sigma ^{1-D}}{\pi nS_{D}r^{n}}\sum_{l=0}^{\infty }
  D_{l}\int_{m}^{\infty }dz\frac{z\Omega _{\alpha \nu _l}(a z,
  b z)}{\sqrt{z^{2}-m^{2}}}G_{\nu _l}^{(\alpha )2}(\alpha z,r z),
\end{equation}
with $\alpha =a,b$, and
\begin{equation}\label{Dlang}
  D_{l}=(2l+D-2)\frac{\Gamma (l+D-2)}{\Gamma (D-1)\,l!}
\end{equation}
being the degeneracy of each angular mode with given $l$, and
$\Gamma (x) $ is the gamma function. Using formula
(\ref{intWeight}), the VEV for the field square can be also
presented in the form
\begin{equation}
\langle 0|\varphi ^{2}(x)|0\rangle =\langle
0_{{\mathrm{m}}}|\varphi ^{2}(x)|0_{{\mathrm{m}}}\rangle +\langle
\varphi ^{2}(x)\rangle _b^{(a)} + \langle \varphi ^{2}(x)\rangle
_b^{(b)}+\langle \varphi ^{2}(x)\rangle ^{(ab)},
\label{fieldsq2sph}
\end{equation}
where $\langle \varphi ^{2}(x)\rangle _b^{(\alpha )}$ is the VEV
induced by a single sphere with radius $\alpha $, the
"interference" term is given by the formula
\begin{eqnarray}
\langle \varphi ^{2}(x)\rangle ^{(ab)} &=&-\frac{\sigma ^{1-D}}{%
\pi nS_{D}br^{n}}\sum_{l=0}^{\infty }D_{l}\int_{m}^{\infty }\frac{zdz}{\sqrt{%
z^{2}-m^{2}}}\frac{\bar{I}_{\nu _{l}}^{(a)}(az)\bar{K}_{\nu _{l}}^{(b)}(bz)}{%
\bar{K}_{\nu _{l}}^{(a)}(az)\bar{I}_{\nu
_{l}}^{(b)}(bz)-\bar{K}_{\nu _{l}}^{(b)}(bz)\bar{I}_{\nu
_{l}}^{(a)}(az)} \nonumber
\\
&&\times \left[ \frac{I_{\nu _{l}}(rz)}{\bar{I} _{\nu
_{l}}^{(b)}(bz)}G_{\nu _{l}}^{(b)}(bz,rz)- \frac{K_{\nu
_{l}}(rz)}{\bar{K} _{\nu _{l}}^{(a)}(az)}G_{\nu _{l}}^{(a)}(az,rz)
\right] , \label{inttermphisq2sph}
\end{eqnarray}
and is finite for all values $a\leq r\leq b$. In the case of the
Dirichlet scalar, by using the relation $I_{\nu }(x)K_{\nu
}(y)>I_{\nu }(y)K_{\nu }(x)$ for $y>x$, it can be seen that $
\langle \varphi ^{2}(x)\rangle ^{(ab)}>0$. Note that for this
boundary condition the both boundary induced terms $ \langle
\varphi ^{2}(x)\rangle _{b}^{(\alpha )} $, $\alpha =a,b$, inside
and outside of a single spherical shell are negative
\cite{Saha03}.

Now let us consider the limiting cases of formula
(\ref{inttermphisq2sph}). In the limit $a\rightarrow 0$ the
subintegrand behaves as $a^{2\nu _l}$, and the dominant
contribution to $\langle \varphi ^{2}(x)\rangle ^{(ab)}$ comes
from the summand with $l=0$. Using the standard formulae for the
Bessel modified functions \cite{abramowiz}, noting that
$B_a=\tilde B_a/a$, and assuming $\nu _0>0$, to the leading order
one has
\begin{eqnarray}
\langle \varphi ^{2}(x)\rangle ^{(ab)} &\approx &\frac{-\sigma ^{1-D}%
}{\pi S_{D}r^{n}\nu _{0}\Gamma ^{2}(\nu _{0})}\frac{n+2\eta _a\nu _{0}}{%
n+2\nu _{0}}\left( \frac{a}{2}\right) ^{2\nu _{0}}\int_{m}^{\infty }%
\frac{z^{2\nu _{0}+1}dz}{\sqrt{z^{2}-m^{2}}}\frac{\bar{K}_{\nu
_{0}}^{(b)}(bz)}{\bar{I}_{\nu _{0}}^{(b)}(bz)}   \nonumber  \\
&&\times I_{\nu _{0}}(rz)\left[ \frac{\bar{K}_{\nu _{0}}^{(b)}(bz)}{\bar{I}%
_{\nu _{0}}^{(b)}(bz)}I_{\nu _{0}}(rz)-2K_{\nu _{0}}(rz)\right]
,\quad a\to 0, \label{deltaphiato0}
\end{eqnarray}
where we use the notation
\begin{equation}
\eta _{\alpha }=\left\{
\begin{array}{rl}
1, & \quad {\rm for}\quad \tilde B_{\alpha }=0 \\
-1, & \quad {\rm for}\quad \tilde B_{\alpha }\neq 0
\end{array}
\right. .  \label{etaalpha}
\end{equation}
For $\nu _0=0$ and $a\to 0$, the dominant contribution behaves as
$\langle \varphi ^{2}(x)\rangle ^{(ab)}\sim 1/\ln a$. Note that
for a minimally coupled scalar ($\xi =0$) one has $\nu _0=D/2-1$.

In the case of a massless scalar the asymptotic behavior of
"interference" part (\ref{inttermphisq2sph}) for large values $b$
and fixed $a$ and $r$ can be obtained by changing the integration
variable to $y=zb$ and expanding the subintegrand in terms of
$a/b$ and $r/b$. The dominant contribution for the summand with a
given $l$ has an order $(a/b)^{2\nu _l}$ (assuming that $A_a\neq
B_a\nu _l$) and the main contribution comes from the $l=0$ term.
The leading term for the corresponding asymptotic expansion can be
presented in the form
\begin{eqnarray}
  \langle \varphi ^2\rangle ^{(ab)}&\approx & -\frac{2\sigma ^{1-D}
  r^{-n}}{\pi S_Da\Gamma ^2(\nu _0+1)}\left( \frac{a}{2b}\right) ^
  {2\nu _0+1}\frac{A_a+B_a\nu _0}{A_a-B_a\nu _0}\left[
  \frac{A_a+B_a\nu _0}{A_a-B_a\nu _0} \left( \frac{a}{r}\right) ^
  {2\nu _0}-2\right] \nonumber \\
  && \times \int_{0}^{\infty }dz \, z^{2\nu _0}\frac{nK_{\nu _0}(z)-
  2\delta _{0\tilde A_b}zK'_{\nu _0}(z)}{nI_{\nu _0}(z)-
  2\delta _{0\tilde A_b}zI'_{\nu _0}(z)}. \label{phi2btoinf}
\end{eqnarray}
Hence, the "interference" part of the VEV of the square of the
field operator vanishes in both limits $a\to 0$ and $b\to \infty
$.

Now we turn to the limit $a,b\rightarrow \infty $ for fixed $b-a$
and $\sigma $. In this limit expression (\ref{inttermphisq2sph})
diverges and the main contribution comes from large values of
$l\sim \sigma [2(1-a/b)]^{-1}$. Introducing in
(\ref{inttermphisq2sph}) a new integration variable $y=bz/\nu
_{l}$, we can replace the Bessel modified functions by their
uniform asymptotic expansions for large values of the order
\cite{abramowiz}. To the leading order this gives
\begin{equation}
\langle \varphi ^{2}(x)\rangle ^{(ab)}\approx -\frac{(b\sigma )^{1-D}}{%
\pi S_D\Gamma (D-1)}\sum_{l=0}^{\infty }l^{D-2}\int_{m_l}^{\infty
}dy \frac{(y^2-m_l^2)^{-1/2}}{c_b(y)e^{2y}/c_a(y)-1} F(y,r) ,
\label{phi2abtoinf1}
\end{equation}
where we have introduced the notations
\begin{equation}\label{ml}
  m_l=(b-a)\sqrt{m^2+(l/b\sigma )^2},\quad c_{\alpha }(y)=
  \frac{\tilde A_{\alpha}+y\tilde B_{\alpha}/(b-a)}{\tilde A_{\alpha}
  -y\tilde B_{\alpha}/(b-a)},\quad \alpha =a,b,
\end{equation}
and
\begin{equation}\label{Fyr}
  F(y,r)=\frac{1}{c_{b}(y)}\exp \left( 2y\frac{r-b}{b-a}\right)
+c_{a}(y)\exp \left( 2y\frac{a-r}{b-a}\right) -2.
\end{equation}
In the limit under consideration in (\ref{phi2abtoinf1}) we can
make the replacement
\begin{equation}\label{abtoinfrepl}
  \sum_{l}l^{D-2}f\left( \frac{l(b-a)}{b\sigma }\right) \to
  \left( \frac{b\sigma }{b-a}\right) ^{D-1}\int_{0}^{\infty }
  dt\, t^{D-2}f(t).
\end{equation}
Further, introducing instead of $y$ a new integration variable
$u=\sqrt{y^2-t^2-m_l^2}$ and converting to polar coordinates on
the plane $(u,t)$, the angular part of the resulting integral is
easily evaluated. Using the standard relations for the gamma
function, one receives
\begin{equation}\label{phi2abtoinf2}
  \langle \varphi ^{2}(x)\rangle ^{(ab)}\approx -
  \frac{(b-a)^{1-D}}{(4\pi )^{D/2}\Gamma (D/2)}\int_{m_0}^{\infty }
  dy\,
  \frac{(y^2-m_0^2)^{D/2-1}}{c_b(y)e^{2y}/c_a(y)-1}
  F(y,r) ,\quad
  a,b\to \infty, b-a={\mathrm{const}}.
\end{equation}
This leading term of the corresponding asymptotic expansion does
not depend on the parameter $\sigma $ and coincides with the
corresponding quantity for two parallel plates with the separation
$b-a$, on background of the Minkowski spacetime.

\begin{figure}[tbph]
\begin{center}
\begin{tabular}{ccc}
\epsfig{figure=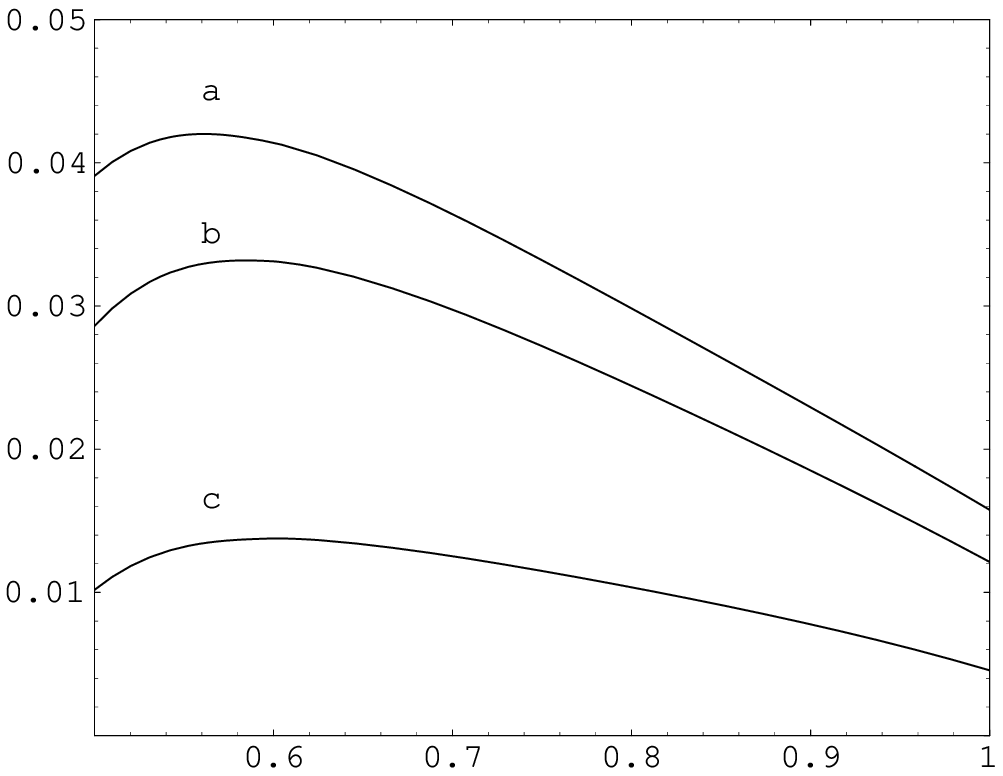,width=6cm,height=5cm} & \hspace*{0.5cm} & %
\epsfig{figure=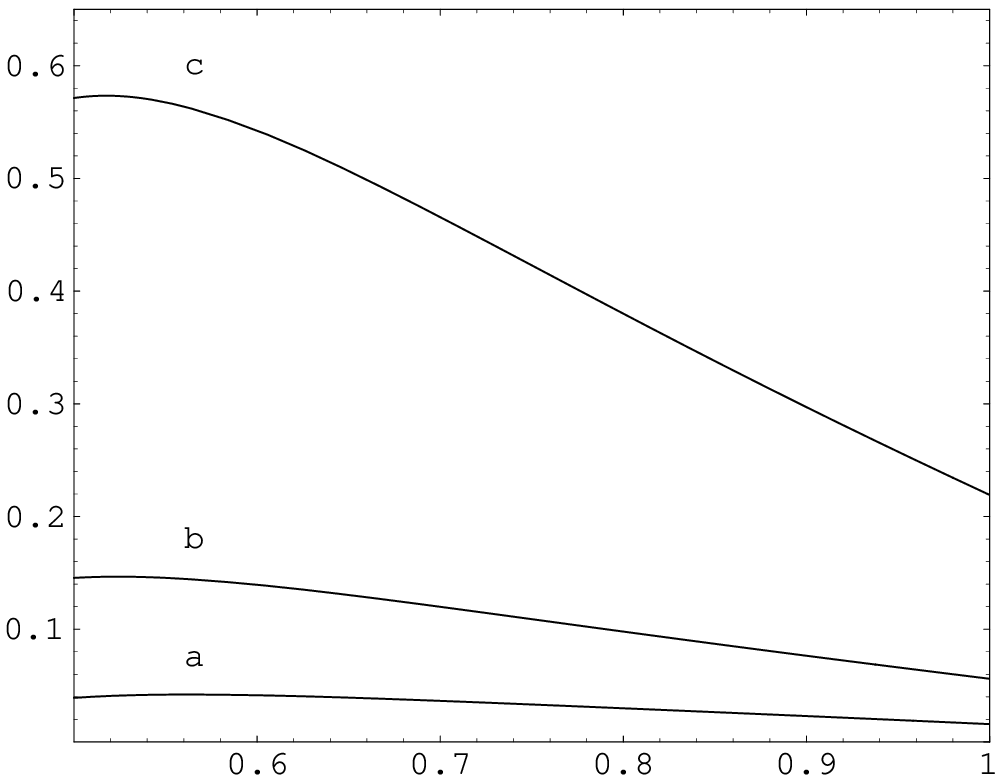,width=6cm,height=5cm}
\end{tabular}
\end{center}
\caption{ The "interference" part $b^{D-1}\langle \varphi
^{2}(x)\rangle ^{(ab)}$ as a function on $r/b$ in the cases of
conformally (left panel) and minimally (right panel) coupled
massless $D=3$ Dirichlet scalars for $a/b=0.5$. The curves are
plotted for $\sigma =1$ (a), $\sigma =0.4$ (b), $\sigma =0.2$
(c).} \label{fig1phi22sph}
\end{figure}

And finally, consider the "interference" term $\langle \varphi ^2
\rangle ^{(ab)}$ in the limit of strong gravitational field,
$\sigma \ll 1$, for fixed values $a,b,r$. For $\xi
>0$, from (\ref{mnulam}) one has $\nu _{l} \gg 1$, and after
introducing in (\ref{inttermphisq2sph}) a new integration variable
$y=z/\nu _{l} $, we can replace the modified Bessel function by
their uniform asymptotic expansions for large values of the order.
The integral over $y$ can be estimated by using the Laplace
method. The main contribution to the sum over $l$ comes from the
summand with $l=0$, and to the leading order we receive
\begin{equation}\label{phi2sigll1}
  \langle \varphi ^2(x)\rangle ^{(ab)}\approx
  \frac{\sigma ^{1-D}\eta _a\eta _b exp[-2\tilde \nu \ln (b/
  a)]}{(2\pi \tilde \nu )^{1/2}r^nS_D\sqrt{b^2-a^2}},\quad
  \tilde \nu =\frac{1}{\sigma }\sqrt{n(n+1)\xi },\quad \sigma \ll 1.
\end{equation}
Note that, using the corresponding expressions for single sphere
parts \cite{Saha03}, we can see that in the limit under
consideration
\begin{equation}\label{phi2sigll12}
  \frac{\langle \varphi ^2(x)\rangle ^{(ab)}}{\langle \varphi ^2(x)
  \rangle ^{(\alpha )}_b}\approx -2\sqrt{\frac{|\alpha ^2-r^2|}{b^2-a^2}}
  \frac{\eta _a\eta _b}{\eta _{\alpha }}exp[-2\tilde \nu |\ln (\alpha
  /r)|], \quad \alpha =a,b,
\end{equation}
and the "interference" term is suppressed compared to the single
sphere contribution. For $\xi =0$ and $\sigma \ll 1$ for the terms
with $l\neq 0$ one has $\nu _{l} \gg 1$. The corresponding
contribution can be estimated by the way similar to that in the
previous case. This contribution is exponentially small. For the
summand with $l=0$ to the leading order over $\sigma $ we have
$\nu _{l} =n/2$ in (\ref{inttermphisq2sph}), and $\langle \varphi
^2 \rangle \sim \sigma ^{1-D}$. As we see, the behavior of the
"interference" part in the VEV of the field square in the limit of
the strong gravitational field is essentially different for
minimally and and non-minimally coupled scalars. This behavior is
clearly seen from figure \ref{fig1phi22sph} where we have plotted
the dependence of the "interference" term $\langle \varphi
^2(x)\rangle ^{(ab)}$ on the ratio $r/b$ in the cases of
conformally ($\xi =\xi _D$, left panel) and minimally ($\xi =0$,
right panel) coupled massless Dirichlet scalars in $D=3$
dimensions for $a/b=0.5$. The separate curves correspond to the
values $\sigma =1$ (a), $\sigma =0.4 $ (b), $\sigma =0.2$ (c).

\section{Vacuum expectation values of the energy-momentum tensor
and the interaction forces between the spheres} \label{secm:4}

In this section we will consider the VEVs for the energy momentum
tensor operator in the region between two spheres on background of
the global monopole spacetime. Using the standard classical
expression, we can write the vacuum energy-momentum tensor as the
coincidence limit
\begin{equation}
\langle 0|T_{ik}(x)|0\rangle =\lim_{x^{\prime }\rightarrow x}\partial
_{i}\partial _{k}^{\prime }\langle 0|\varphi (x)\varphi (x^{\prime
})|0\rangle +\left[ \left( \xi -\frac{1}{4}\right) g_{ik}\nabla _{l}\nabla
^{l}-\xi \nabla _{i}\nabla _{k}-\xi R_{ik}\right] \langle 0|\varphi
^{2}(x)|0\rangle ,  \label{mvevEMT}
\end{equation}
where for the point-like global monopole spacetime the nonzero
components of the Ricci tensor are given by expressions
\begin{equation}
R_{2}^{2}=R_{3}^{3}=\cdots =R_{D}^{D}=n\frac{\sigma ^{2}-1}{\sigma
^{2}r^{2}},  \label{mRictens}
\end{equation}
with the indices $2,3,\ldots ,D$ corresponding to the coordinates
$\theta _{1},\theta _{2},\ldots ,\phi $ respectively. Substituting
the Wightman function (\ref{regWightab1}) into (\ref{mvevEMT}), we
obtain that the vacuum energy-momentum tensor has the diagonal
form (as expected by the symmetry of the model)
\begin{equation}
\langle 0|T_{i}^{k}|0\rangle ={\rm diag}\left( \varepsilon ,-p,-p_{\perp
},\ldots ,-p_{\perp }\right) ,  \label{diagEMT}
\end{equation}
where the vacuum energy density $\varepsilon $ and the effective
pressures in radial, $p$, and azimuthal, $p_{\perp }$, directions
are functions of the radial coordinate only. Using the Wightman
function from (\ref{regWightab1}) and the VEV for the field square
from (\ref{fieldsqform1}), the components of the vacuum
energy-momentum tensor can be presented in the form
\begin{equation}
q(a,b,r)=q(a,r)+q_{a}(a,b,r),\quad a<r<b,\quad q=\varepsilon ,p,p_{\perp },
\label{compab1}
\end{equation}
where $q(a,r)$ are the corresponding functions for the vacuum
outside a single sphere with radius $a$. In (\ref{compab1}) the
additional components are in the form
\begin{eqnarray}
q_{a}(a,b,r)&=&-\frac{\sigma ^{1-D}}{2\pi
r^{n}S_{D}}\sum_{l=0}^{\infty }D_{l}\int_{m}^{\infty
}dz\frac{z^{3}\Omega _{a\nu _{l}}(az,b
z)}{\sqrt{z^{2}-m^{2}}}F_{\nu _{l}}^{(q)}\left[ G_{\nu
_{l}}^{(a)}(az,y),G_{\nu _{l}}^{(a)}(az,y)\right] _{y=zr} ,
 \label{q1ab} \\
 q&=&\varepsilon ,\,p,\,p_{\perp }, \nonumber
\end{eqnarray}
where for arbitrary functions $f(y)$ and $g(y)$ the functions
$F_{\nu }^{(q)}\left[ f(y),g(y)\right] $ are defined by the
relations
\begin{eqnarray}
F_{\nu _{l}}^{(\varepsilon )}\left[ f(y),g(y)\right]  &=&(1-4\xi
)\left[ f^{\prime }g^{\prime }-\frac{n}{2y}(f g)^{\prime }+\left(
\frac{\nu _{l}^{2}}{ y^{2}}-\frac{1+4\xi -2(mr/y)^{2}}{1-4\xi
}\right) f g \right]
\label{Fineps} \\
F_{\nu _{l}}^{(p)}\left[ f(y),g(y)\right]  &=&f^{\prime }g^{\prime
}+\frac{\tilde{\xi} }{2y}(f g)^{\prime }-\left( 1+\frac{\nu
_{l}^{2}+\tilde{\xi}n/2}{y^{2}}
\right) f g,\;\tilde{\xi}=4(n+1)\xi -n  \label{Finperad} \\
F_{\nu _{l}}^{(p_{\perp })}\left[ f(y),g(y)\right]  &=&(4\xi
-1)f^{\prime }g^{\prime }-\frac{\tilde{\xi}}{2y}(f g)^{\prime
}+\left[ 4\xi -1+\frac{\nu
_{l}^{2}(1+\tilde{\xi})+\tilde{\xi}n/2}{(n+1)y^{2}}\right] f g ,
\label{Finpeaz}
\end{eqnarray}
and the prime denotes the differentiation with respect to $y$.
Quantities (\ref{q1ab}) are finite for $a\leq r<b$ and diverge at
the surface $r=b$. Similar to the Wightman function, the
components of the vacuum energy-momentum tensor for a single
sphere case can be presented in the form
\begin{equation}\label{qar1sph}
  q(a,r)=q_{{\mathrm{m}}}(r)+q_b(a,r),
\end{equation}
where $q_{{\mathrm{m}}}(r)$ are the corresponding quantities for
the boundary-free monopole geometry and the expressions for the
sphere induced parts $q_b(a,r)$ are given in \cite{Saha03}.

On the basis of formula (\ref{regWightab2}), the vacuum
energy-momentum tensor components may be written in another
equivalent form:
\begin{equation}
q(a,b,r)=q(b,r)+q_{b}(a,b,r),\quad a<r<b,\quad q=\varepsilon ,p,p_{\perp },
\label{compab2}
\end{equation}
with $q(b,r)$ being the corresponding components for the vacuum inside a
single sphere with radius $b$. Here the additional components are given by
the formula
\begin{eqnarray}
q_{b}(a,b,r)&=&-\frac{\sigma ^{1-D}}{2\pi
r^{n}S_{D}}\sum_{l=0}^{\infty }D_{l}\int_{m}^{\infty
}dz\frac{z^{3}\Omega _{b\nu _{l}}(az,b
z)}{\sqrt{z^{2}-m^{2}}}F_{\nu _{l}}^{(q)}\left[ G_{\nu
_{l}}^{(b)}(b z,y),G_{\nu _{l}}^{(b)}(b z,y)\right] _{y=zr} ,
\label{q2ab} \\
q&=&\varepsilon ,\,p,\,p_{\perp }.
\end{eqnarray}
This expressions are finite for all $a<r\leq b$ and diverge at the inner
sphere surface $r=a$.

It follows from the above that if we present the vacuum
energy-momentum tensor components in the form
\begin{equation}
q(a,b,r)=q_{{\mathrm{m}}}(r)+q_b(a,r)+q_b(b,r)+\Delta
q(a,b,r),\quad a<r<b, \label{qinterf}
\end{equation}
then the quantities
\begin{equation}
\Delta q(a,b,r)=q_{a}(a,b,r)-q(b,r)=q_{b}(a,b,r)-q(a,r)
\label{deltaq0}
\end{equation}
are finite for all $a\leq r\leq b$. Near the surface $r=a$ it is
suitable to use the first equality in (\ref{deltaq0}), as for $
r\rightarrow a$ both summands are finite. For the same reason the
second equality is suitable for calculations near the outer
surface $r=b$. Using formula (\ref{intWeightrr}) for the
corresponding part of the Wightman function, it can be seen that
the following formula takes place for the "interference" parts
\begin{eqnarray}
\Delta q(a,b,r)&=&\frac{\sigma ^{1-D}}{2\pi
r^{n}S_{D}}\sum_{l=0}^{\infty }D_{l}\int_{m}^{\infty
}\frac{z^{3}dz}{\sqrt{z^{2}-m^{2}}}\frac{\bar{I}_{\nu _l
}^{(a)}(az)\bar{K}_{\nu _l}^{(b)}(b z)}{\bar{I}_{\nu
_l}^{(a)}(az)\bar{K}_{\nu _l }^{(b)}(b z)-\bar{I}_{\nu _l}^{(b)}(b
z)\bar{K}_{\nu _l}^{(a)}(az)} \nonumber \\
&& \times \left\{ \frac{F^{(q)}_{\nu _l}[I_{\nu }(y),G_{\nu
_l}^{(b)}(b z,y)]}{\bar{I}_{\nu _l}^{(b)}(b z)}
-\frac{F^{(q)}_{\nu _l}[K_{\nu }(y),G_{\nu _l}^{(a)}(a
z,y)]}{\bar{K}_{\nu _l}^{(a)}(az)}\right\} _{y=zr}. \label{deltaq}
\end{eqnarray}
It can be checked that this quantities satisfy the covariant
continuity equation
\begin{equation}\label{covconteq}
  r\frac{d\Delta p}{dr}+(D-1)(\Delta p-\Delta p_{\perp })=0.
\end{equation}
Note that the ambiguities of the renormalization procedure for the
VEV of the energy-momentum tensor in the form of an arbitrary mass
scale (see \cite{Mazz91}) are contained in the boundary-free parts
$q_{{\mathrm{m}}}(r)$ of the corresponding components. The
boundary induced parts $q_b(\alpha ,r)$, $\alpha =a,b$ and $\Delta
q(a,b,r) $ are unambiguously defined for $a<r<b$. In particular,
for the massless conformally coupled scalar they contain no
conformal anomalies and are traceless.

Now we turn to the investigation of limiting cases of formula
(\ref{deltaq}). In the limit $a \to 0$, the subintegrand behaves
as $a^{2\nu _l}$ and the dominant contribution comes from the
$l=0$ term:
\begin{eqnarray}
\Delta q(a,b,r) &\approx &-\frac{\sigma ^{1-D} (a/2b)^{2\nu
_0}}{2\pi S_{D}r^{n}b^3\nu _{0}\Gamma ^{2}(\nu _{0})}\frac{n+2\eta
_a\nu _{0}}{ n+2\nu _{0}}\int_{mb}^{\infty } \frac{z^{2\nu
_{0}+3}dz}{\sqrt{z^{2}-m^{2}b^2}}\frac{\bar{K}_{\nu
_{0}}^{(b)}(z)}{\bar{I}_{\nu _{0}}^{(b)}(z)}   \nonumber  \\
&&\times \left\{ \frac{\bar{K}_{\nu _{0}}^{(b)}(z)}{\bar{I} _{\nu
_{0}}^{(b)}(z)}F_{\nu _0}^{q}[I_{\nu _{0}}(y),I_{\nu
_{0}}(y)]-2F_{\nu _0}^{q}[I_{\nu _{0}}(y),K_{\nu _{0}}(y)]\right\}
_{y=zr/b} ,\quad a\to 0, \label{deltaqato0}
\end{eqnarray}
assuming that $\nu _0>0$. For $\nu _0=0$ and $a\to 0$ the
"interference" parts (\ref{deltaq}) behave as $1/\ln a$. Consider
the limit $b\to \infty $ for fixed $a$ and $r$ in the case of a
massless scalar field. By changing the integration variable to
$y=bz$ and using the formula for the modified Bessel functions in
the case of small values of the argument, we see that to the
leading order the subintegrand in (\ref{deltaq}) behaves as
$(a/b)^{2\nu _l+1}$ (assuming that $\nu _l>0$). The leading
contribution is due to the $l=0$ term and one has:
\begin{eqnarray}
  \Delta q(a,b,r) &\approx & -\frac{\sigma ^{1-D}(a/2b)^
  {2\nu _0+1}
  }{\pi S_Dar^{D}\Gamma ^2(\nu _0+1)}\frac{A_a+B_a\nu _0}{A_a-B_a\nu _0}
  \left[ (2\nu _0+n)f_{1\nu _0}^{(q)}
  \frac{A_a+B_a\nu _0}{A_a-B_a\nu _0} \left( \frac{a}{r}\right) ^
  {2\nu _0}+2f_{2\nu _0}^{(q)}\right] \nonumber \\
  && \times \int_{0}^{\infty }dz \, z^{2\nu _0}\frac{nK_{\nu _0}(z)-
  2\delta _{0\tilde A_b}zK^{\prime }_{\nu _0}(z)}{nI_{\nu _0}(z)-
  2\delta _{0\tilde A_b}zI^{\prime }_{\nu _0}(z)},\quad b\to
  \infty ,
  \label{deltaqbtoinf}
\end{eqnarray}
with notations
\begin{eqnarray}\label{f1nu0}
&& f_{1\nu }^{(\varepsilon )}=\nu (1-4\xi ),\quad f_{1\nu
}^{(p)}=-\frac{\tilde \xi }{2},\quad f_{1\nu }^{(p_{\perp
})}=\frac{\nu +1/2}{n+1}\tilde \xi ,\\
&& f_{2\nu }^{(\varepsilon )}=0,\quad f_{2\nu }^{(p)}=-n-1,\quad
f_{2\nu }^{(p_{\perp })}=4n(n+1)\xi /\sigma ^2 .
\end{eqnarray}
As we see in both limits $a\to 0$ and $b\to \infty $ the
"interference" parts for the energy-momentum tensor components
vanish.

In the limit $a,b\to \infty $ for fixed values $b-a$ and $\sigma
$, by the calculations similar to those for the "interference"
part of the field square, one receives
\begin{equation}\label{deltaqabtoinf}
  \Delta q(a,b,r) \approx -\frac{(b-a)^{-D-1}}{(4\pi )^{D/2}
  \Gamma (D/2+1)}\int_{m_0}^{\infty }dy \frac{(y^2-
  m_0^2)^{D/2}}{c_b(y)e^{2y}/c_a(y)-1}\Delta F^{q}(y,r),
\end{equation}
where
\begin{equation}\label{deltaFyr}
  \Delta F^{\varepsilon }(y,r)=-\Delta F^{p_{\perp } }(y,r)=1-
  \frac{4D(\xi -\xi _D)y^2-m_0^2}{2(y^2-m_0^2)}[F(y,r)+2],\quad
  \Delta F^{p}=1,
\end{equation}
with the function $F(y,r)$ defined by (\ref{Fyr}). These
expressions are exactly the same as the corresponding expressions
for the geometry of two parallel plates on the Minkowski
background investigated in \cite{Rome02} (note that in
\cite{Rome02}, the notations $\tilde B_a/\tilde A_a=\beta _1$ and
$\tilde B_b/\tilde A_b=-\beta _2$ are used) for a massless scalar
and in Ref. \cite{Mate03} for the massive case.

And finally, let us consider the limit of the strong gravitational
field, corresponding to $\sigma \ll 1$. In this limit for $\xi >0$
one has $\nu _l\gg 1$. After the change of variable to $y=bz/\nu
_l $, we replace in (\ref{deltaq}) the Bessel modified functions
by their uniform asymptotic expansions for large values of the
order. The leading contribution comes from the $l=0$ summand and
we have the following limit for the "interference" parts of the
vacuum energy-momentum components:
\begin{equation}\label{deltaqsigll1}
\Delta p\approx -\frac{\Delta p_{\perp }}{D-1}\approx
-\frac{\sigma ^{1-D}\eta _a\eta _b\tilde \nu
^{3/2}}{r^DS_D\sqrt{2\pi (b^2-a^2)}}e^{-2\tilde \nu \ln (b/a)},
\quad \Delta \varepsilon /\Delta p\sim \sigma ,
\end{equation}
where $\tilde \nu $ is defined in (\ref{phi2sigll1}). In this
limit the "interference" parts are exponentially suppressed with
respect to single spheres contributions.

Now we turn to the interaction forces between the spheres. The
vacuum force acting per unit surface of the sphere $r=\alpha $,
$\alpha =a,b$, is determined by the ${}^{1}_{1}$--component of the
vacuum energy-momentum tensor at this point. By virtue of
relations (\ref{compab1}) and (\ref{compab2}), the corresponding
effective pressures can be presented as a sum of the pressure for
a single sphere with $r =\alpha$ when the second sphere is absent
and the pressure induced by the presence of the second sphere,
$p_{\alpha }(a,b,r=\alpha )$. The first term is divergent due to
the well-known surface divergences and needs additional
regularization. The second term is finite and can be termed as an
interaction force between the spheres. This additional radial
vacuum pressure on the sphere with $r=\alpha $, $\alpha =a,b$ due
to the existence of the second sphere can be found from
(\ref{q1ab}) and (\ref{q2ab}), respectively. Using the relations
\begin{equation}
G_{\nu }^{(r)}(rz,rz)=-B_{r},\quad rz\frac{\partial }{\partial y}G_{\nu
}^{(r)}(rz,y)\mid _{y=rz}=A_{r},\quad r=a,b,  \label{Grelab}
\end{equation}
they can be presented in the form
\begin{eqnarray}
p_{\alpha }(a,b,r=\alpha ) &=&-\frac{\sigma ^{1-D}}{2\pi \alpha ^{D}S_{D}}%
\sum_{l=0}^{\infty }D_{l}\int_{m}^{\infty }\frac{zdz}{\sqrt{z^{2}-m^{2}}}%
\Omega _{\alpha \nu _{l}}(az,bz)  \label{paabb} \\
&\times &\left\{ \tilde{A}_{\alpha }^{2}-4(D-1)\xi \tilde{A}_{\alpha
}B_{\alpha }-\left[ z^{2}\alpha ^{2}+\nu _{l}^{2}-n^{2}/4\right] B_{\alpha
}^{2}\right\} ,\qquad \alpha =a,b.  \nonumber
\end{eqnarray}
Unlike the self-action forces, these quantities are finite for
$a<b$ and need no further regularization. For the Dirichlet scalar
one has $\Omega _{\alpha \nu _{l}}(az,bz)>0$ and, hence,
$p_{\alpha }(a,b,r=\alpha ) <0$. This means that in this case the
vacuum interaction forces between the spheres are attractive.
Using the Wronskian relation for the Bessel modified functions, it
can be seen that
\begin{equation}
\left[ A_{\alpha }-B_{\alpha }(z^{2}\alpha ^{2}+\nu ^{2})\right] \Omega
_{\alpha \nu }(az,bz)=-n_{\alpha }\alpha \frac{\partial }{\partial \alpha }%
\ln \left| 1-\frac{\bar{I}_{\nu }^{(a)}(az)}{\bar{I}_{\nu }^{(b)}(bz)}\frac{%
\bar{K}_{\nu }^{(b)}(bz)}{\bar{K}_{\nu }^{(a)}(az)}\right| ,
\label{Omeganew}
\end{equation}
where $n_{a}=1,\,n_{b}=-1$. This allows us to write the
expressions (\ref{paabb}) for the interaction forces per unit
surface in another equivalent form:
\begin{eqnarray}
p_{\alpha }(a,b,r=\alpha ) &=&\frac{n_{\alpha }\sigma ^{1-D}}{2\pi \alpha
^{D-1}S_{D}}\sum_{l=0}^{\infty }D_{l}\int_{m}^{\infty }\frac{z dz}{\sqrt{%
z^{2}-m^{2}}}\left[ 1-\frac{\tilde \xi \tilde{A}_{\alpha }B_{\alpha }}{%
A_{\alpha }^{2}-B_{\alpha }^{2}(z^{2}\alpha ^{2}+\nu _{l}^{2})}\right]
\label{paabb1} \\
&&\times \frac{\partial }{\partial \alpha }\ln \left|
1-\frac{\bar{I}_{\nu _{l}}^{(a)}(az)}{\bar{I}_{\nu _{l}}^{(b)}(b
z)}\frac{\bar{K}_{\nu _{l}}^{(b)}(b z)}{\bar{K}_{\nu
_{l}}^{(a)}(az)}\right| ,\qquad \alpha =a,b, \nonumber
\end{eqnarray}
where $\tilde \xi $ is defined in (\ref{Finperad}). For Dirichlet
and Neumann scalars the second term in the square brackets is
zero.

Let us consider the interaction forces between the spheres in
limiting cases. For small values of the radius of the inner
sphere, $a\to 0$, the leading contribution comes from the $l=0$
summand. The corresponding terms are given by formulae
\begin{eqnarray}\label{paato0}
  p_a(a,b,r=a)&\approx & -\frac{\sigma ^{1-D}a^{2\nu _0-D}}{2^{2\nu _0-1}
  \Gamma ^2(\nu _0)\pi S_D}\frac{n+2\eta _a\nu _0}{n+2\nu _0}
  \int_{m}^{\infty }dz\frac{z^{2\nu _0+1}}{\sqrt{z^2-m^2}}\frac{\bar K_{
  \nu _0}^{(b)}(b z)}{\bar I_{\nu _0}^{(b)}(b z)}, \\
p_b(a,b,r=b)&\approx & -\frac{\sigma ^{1-D}a^{2\nu _0}}{2^{2\nu
_0}\nu _0 \Gamma ^2(\nu _0)\pi S_Db^D}\frac{n+2\eta _a\nu
_0}{n+2\nu _0}
  \int_{m}^{\infty }dz\frac{z^{2\nu _0+1}}{\sqrt{z^2-m^2}
  \bar I_{\nu _0}^{(b)2}(b z)} \nonumber \\
  && \times \left\{ \tilde A^2_b-4(D-1)\xi \tilde A_bB_b-\left[
  z^2b^2+(1/\sigma ^2-1)n(n+1)\xi \right] B_b^2\right\} ,\label{pbbto0}
\end{eqnarray}
where $\eta _a$ is defined in (\ref{etaalpha}).

For large values of the radius of the outer sphere, $b\to \infty
$, in the case of a massless scalar to the leading order we find
\begin{eqnarray}
p_{a}(a,b,r=a)&\approx &-\frac{2\sigma ^{1-D}(A_{a}-B_{a}\nu _{0})^{-2}}{%
\pi a^{D}bS_{D}\Gamma ^{2}(\nu _{0})}\left( \frac{a}{2b}\right) ^{2\nu _{0}}%
\left[ \tilde A_{a}^{2}-4(D-1)\xi \tilde A_{a}B_{a}-(\nu
_{0}^{2}-n^{2}/4)B_{a}^{2}\right]
\nonumber  \\
&&\times \int_{0}^{\infty }dz\,z^{2\nu _{0}}\frac{nK_{\nu
_{0}}(z)-2\delta _{0\tilde A_{b}}K_{\nu _{0}}^{\prime
}(z)}{nI_{\nu _{0}}(z)-2\delta _{0\tilde A_{b}}I_{\nu
_{0}}^{\prime }(z)},  \label{paabtoinf} \\
p_{b}(a,b,r=b)&\approx &
-\frac{\sigma ^{1-D}(a/2b)^{2\nu _{0}}}{\pi \nu
_{0}\Gamma ^{2}(\nu _{0})b^{D+1}S_{D}}\frac{A_{a}+B_{a}\nu _{0}}{%
A_{a}-B_{a}\nu _{0}} \int_{0}^{\infty }dz\,\frac{z^{2\nu
_{0}}}{I_{\nu _{0}}^{2}(z)}.  \label{pbbbtoinf}
\end{eqnarray}
In (\ref{pbbbtoinf}) we have assumed that $\tilde A_b\neq 0$. In
the case $\tilde A_b =0 $ (Neumann boundary condition on the outer
sphere) the subintegrand is equal to $-(z^2+\nu _0^2-n^2/4)z^{2\nu
_0}[nI_{\nu _0}(z)/2-zI^{\prime }_{\nu _0}(z)]^{-2}$.

Now let us consider the limit $\sigma \ll 1$. For $\xi > 0$ one
has $\nu _l\gg 1$ for all $l$, and to the leading order we have
\begin{equation}\label{palphsigll1}
  p_{\alpha }(a,b,r=\alpha )\approx -\frac{\eta _{a}\eta _{b}\sigma ^{1-D}
  \tilde \nu ^{3/2}}{\sqrt{2\pi }\alpha ^{D}S_{D}\sqrt{b^{2}-a^{2}}}e^{-2
  \tilde \nu \ln (b/a)},\quad \alpha =a,b.
\end{equation}
In the case of a minimally coupled scalar, $\xi =0$, the main
contribution is due to the summand $l=0$ with $\nu _l=n/2$ and is
of an order $\sigma ^{1-D}$. In this case the contributions coming
from the $l>0$ summands are exponentially suppressed.

\begin{figure}[tbph]
\begin{center}
\begin{tabular}{ccc}
\epsfig{figure=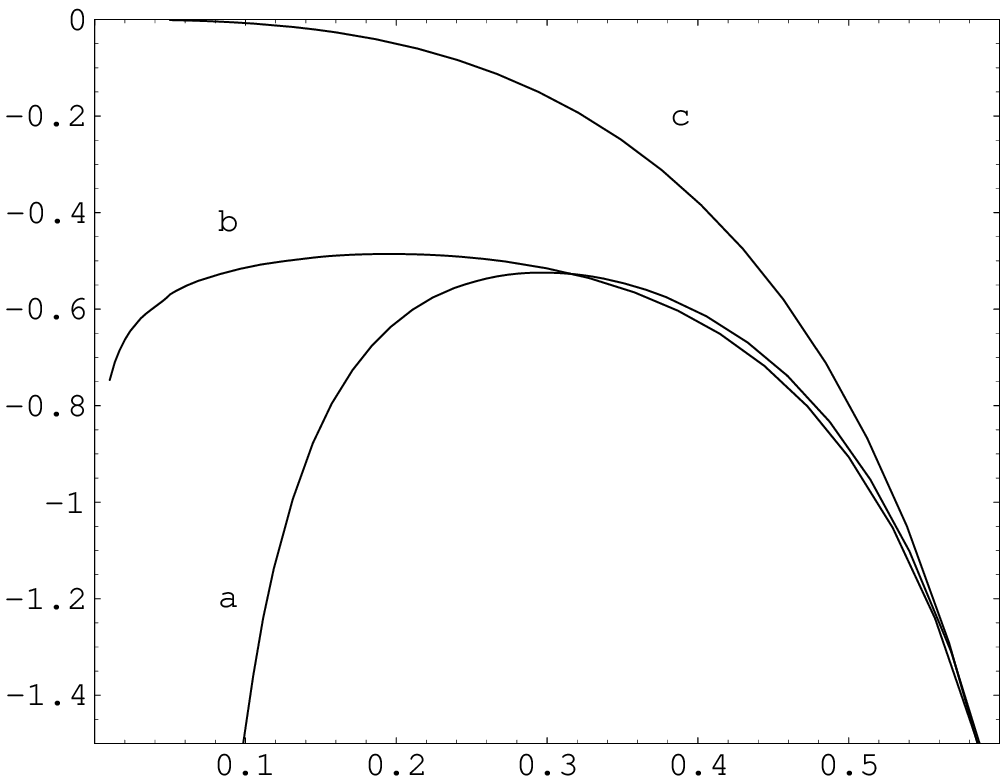,width=6cm,height=5cm} & \hspace*{0.5cm} & %
\epsfig{figure=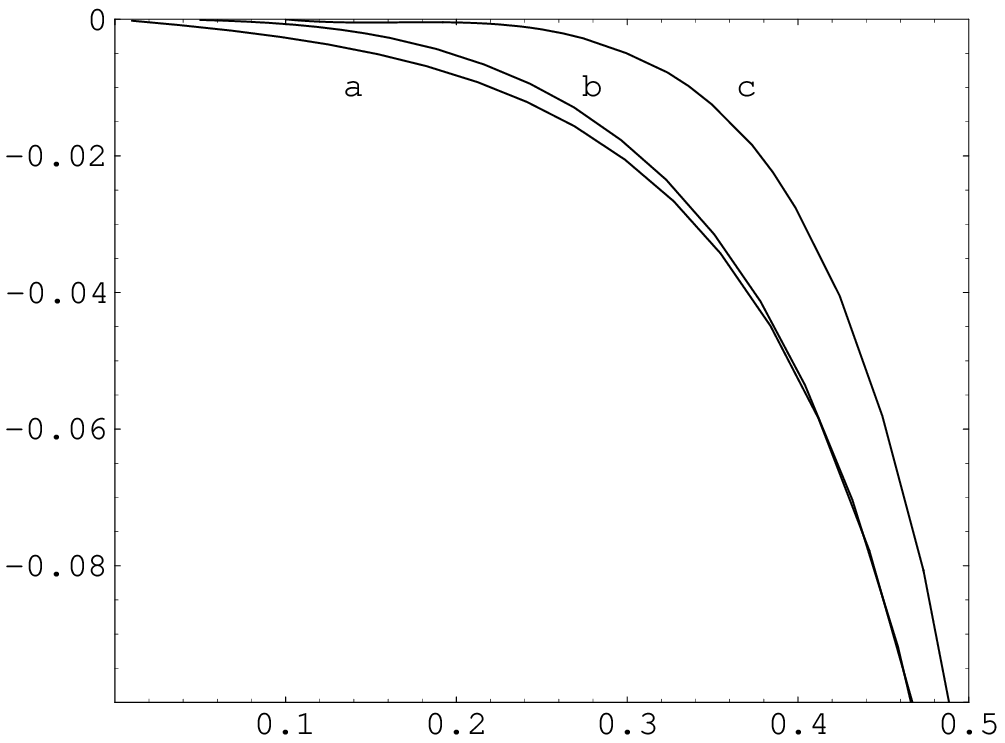,width=6cm,height=5cm}
\end{tabular}
\end{center}
\caption{ The interaction forces between the spheres,
$b^{D-1}p_{a}(a,b,r=a)$ (left panel) and $b^{D-1}p_{b}(a,b,r=b)$,
(right panel) as functions on $a/b$ in the case of a conformally
coupled massless $D=3$ Dirichlet scalar on background of the
global monopole geometry with $\sigma =1$ (a), $\sigma =0.4$ (b),
$\sigma =0.2$ (c).} \label{fig2intforce}
\end{figure}

\begin{figure}[tbph]
\begin{center}
\begin{tabular}{ccc}
\epsfig{figure=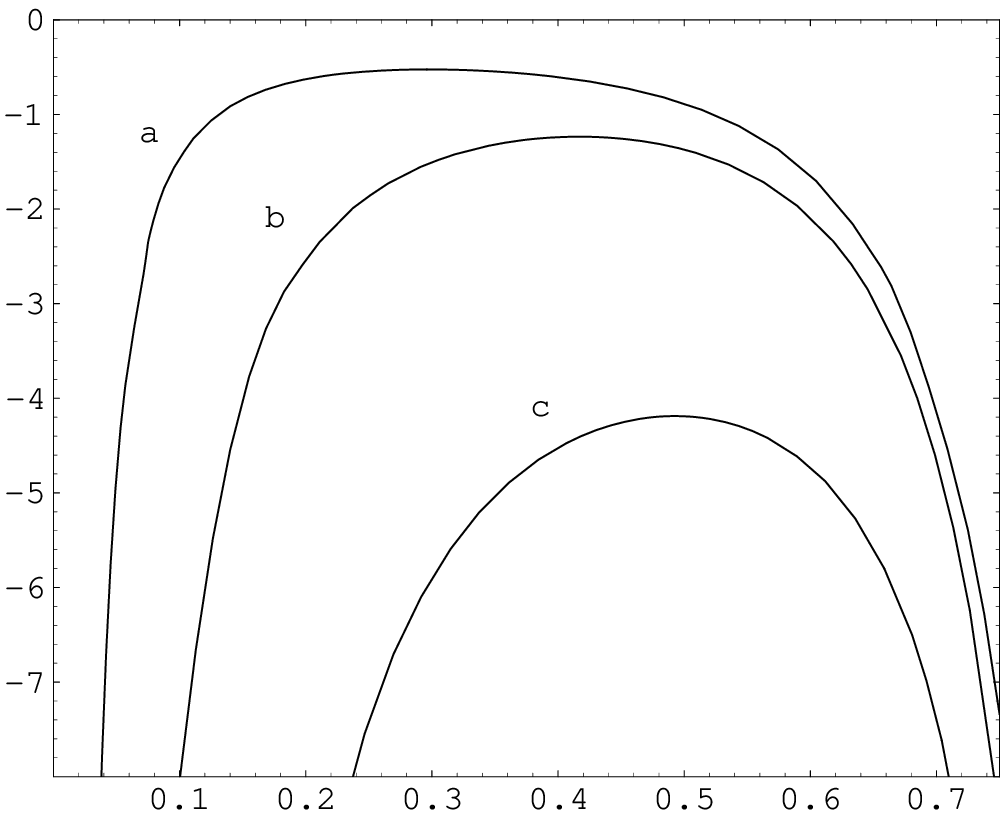,width=6cm,height=5cm} & \hspace*{0.5cm} & %
\epsfig{figure=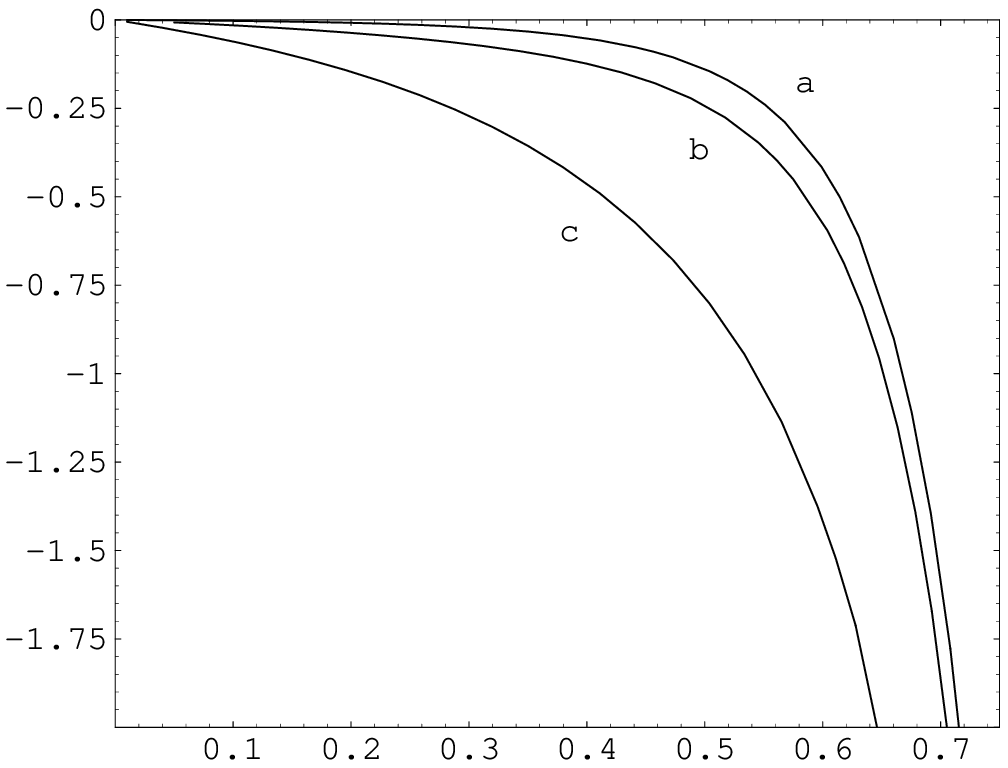,width=6cm,height=5cm}
\end{tabular}
\end{center}
\caption{ The interaction forces between the spheres,
$b^{D-1}p_{a}(a,b,r=a)$ (left panel) and $b^{D-1}p_{b}(a,b,r=b)$,
(right panel) as functions on $a/b$ in the case of a minimally
coupled massless $D=3$ Dirichlet scalar on background of the
global monopole geometry with $\sigma =1$ (a), $\sigma =0.4$ (b),
$\sigma =0.2$ (c).} \label{fig3intforce}
\end{figure}

As in the cases of the VEVs for the field square and
energy-momentum tensor, the dependence of the interaction forces
on the parameter $\sigma $ is essentially different for minimally
and non-minimally coupled scalars. This feature is illustrated in
figures \ref{fig2intforce} and \ref{fig3intforce} where we have
plotted the dependence of the interaction forces $p_{a}(a,b,r=a)$
(left panels) and $p_{b }(a,b,r=b)$ (right panels) between the
spheres on $a/b$ for conformally and minimally coupled $D=3 $
Dirichlet scalars on background of the global monopole background
with $\sigma =1$ (a), $\sigma =0.4$ (b), $\sigma =0.2$ (c).

\section{Conclusion} \label{secm:5}

In this paper, we present the quantum vacuum effects produced by
two concentric spherical shells in the $D+1$-dimensional
point-like global monopole spacetime defined by the line element
(\ref{mmetric}). The case of a massive scalar field with general
curvature coupling parameter and satisfying the Robin boundary
conditions on the spheres is considered. To derive formulae for
the vacuum expectation values of the square of the field operator
and the energy-momentum tensor, we first construct the positive
frequency Wightmann function. This function is also important in
considerations of the response of a particle detector at a given
state of motion through the vacuum under consideration
\cite{Birr82}. The application of the generalized Abel-Plana
formula to the mode sum over zeros of the combinations of the
cylinder functions allows us to extract the part due to a single
sphere on background of the global monopole geometry, formula
(\ref{intWeight}). In this formula the "interference" part $\Delta
W(x,x')$, given by formula (\ref{intWeightrr}), is finite in the
coincidence limit for all values $a\leq r\leq b$. The VEV of the
square of the scalar field, $\langle 0|\varphi ^2(x)|0\rangle $,
is given by the evaluation of the Wightman function in the
coincidence limit. The expectation values for the energy-momentum
tensor are obtained by applying on the corresponding Wightman
function a certain second-order differential operator and taking
the coincidence limit. In both cases the expectation values can be
presented as a sum of boundary-free global monopole, single sphere
induced and "interference" terms. The VEVs related to the
spacetime of a point-like global monopole without boundaries are
considered in \cite{Hisc90,Mazz91,Beze99,Beze02}, and the effects
produced by a single sphere are investigated in a previous paper
\cite{Saha03}. Note that for the points not lying on the spheres,
the boundary induced terms are unambiguously defined and the
ambiguities in the renormalization procedure in the form of an
arbitrary mass scale are contained in the boundary-free parts
only. In particular, for the massless conformally coupled scalar
the boundary induced vacuum energy-momentum tensor contains no
conformal anomalies and is traceless. In this paper we concentrate
on the "interference" parts in the local Casimir densities and
interaction forces between the spheres. The application of the
generalized Abel-Plana formula allows us to derive closed
expressions for these quantities, given by formula
(\ref{inttermphisq2sph}) in the case of the field square, and by
formula (\ref{deltaq}) for the energy-momentum tensor. As bulk
divergences are contained in the part corresponding to the global
monopole geometry without boundaries, and the surface divergences
are contained in the single sphere parts, the "interference" parts
are finite for all values $a\leq r\leq b$. In particular, the
integrals in the corresponding formulae are exponentially
convergent and they are useful for numerical evaluations. We have
considered various limiting cases of the formulae for the
"interference" parts. In the limits $a\to 0$ or $b\to \infty $ for
a fixed value of $r$, these parts vanish as $a^{2\nu _0}$ and
$b^{-(2\nu _0+1)}$ respectively, where $\nu _0$ is defined by
relation (\ref{mnulam}) with $l=0$. In the limit $a,b\to \infty $
for a fixed value $b-a$, the leading terms in the boundary
produced parts do not depend on the parameter $\sigma $ associated
with the solid angle deficit, and are exactly the same as the
corresponding quantities for the geometry of two parallel Robin
plates on the Minkowski background. We have also investigated the
limit of strong gravitational fields corresponding to small values
of the parameter $\sigma $, $\sigma \ll 1$. In this limit the
behaviour of the Casimir densities is drastically different for
minimally ($\xi =0$) and non-minimally ($\xi \neq 0$) coupled
scalars. In the minimal coupling case the leading terms of the
corresponding asymptotic expansions for both field square and the
energy-momentum tensor VEVs behave as $\sigma ^{1-D} $. For a
non-minimally coupled scalar, the "interference" parts behave as
$\langle \varphi ^2 \rangle ^{(ab)}\sim \sigma ^{3/2-D}\exp
(-\gamma /\sigma )$ and $\Delta p\sim \Delta p_{\perp } \sim
\Delta \varepsilon /\sigma \sim \sigma ^{-D-1/2}\exp (-\gamma
/\sigma )$, with $\gamma =2\sqrt{n(n+1)\xi }\ln (b/a)$. Note that
in this case the "interference" parts are exponentially suppressed
with respect to single sphere contributions. The vacuum forces
acting on spheres contain two terms. The first ones are the forces
acting on a single sphere then the second boundary is absent. Due
to the well--known surface divergences in the VEV's of the
energy-momentum tensor these forces are infinite and need an
additional regularization. The another terms in the vacuum forces
are finite and are induced by the presence of the second boundary.
They correspond to the interaction forces per unit surface between
the spheres and are determined by formula (\ref{paabb}). For the
Dirichlet scalar these forces are always attractive. In the limit
of the strong gravitational field, $\sigma \ll 1$, for the
minimally coupled scalar field the interactions forces behave as
$\sigma ^{1-D}$, whereas for a non-minimally coupled scalar they
are exponentially small, $p_{\alpha }(a,b,r=\alpha )\sim \sigma
^{-D-1/2}\exp (-\gamma /\sigma )$.

\section*{Acknowledgement }

We acknowledge support from the Research Project of the Kurdistan
University. The work of AAS was supported in part by the Armenian
Ministry of Education and Science (Grant No.~0887).

\end{document}